\documentclass[preprint,aps]{revtex4}
\input{epsf}
\begin{document}
\title{Properties of a Dilute Bose Gas near a Feshbach Resonance}
\author{Lan Yin}
\email{yinlan@pku.edu.cn} \altaffiliation[Former
address:]{Department of Physics, University of Washington,
Seattle, WA 98195-1560} \author{Zhen-Hua Ning}\affiliation{School
of Physics, Peking University, Beijing 100871, P. R. China}
\date{\today}

\begin{abstract}
In this paper, properties of a homogeneous Bose gas with a
Feshbach resonance are studied in the dilute region at zero
temperature. The stationary state contains condensations of atoms
and molecules.  The ratio of the molecule density to the atom
density is $\pi na^3$.  There are two types of excitations,
molecular excitations and atomic excitations.  Atomic excitations
are gapless, consistent with the traditional theory of a dilute
Bose gas.  The molecular excitation energy is finite in the long
wavelength limit as observed in recent experiments on $^{85}$Rb.
In addition, the decay process of the condensate is studied.  The
coefficient of the three-body recombination rate is about 140
times larger than that of a Bose gas without a Feshbach resonance,
in reasonably good agreement with the experiment on $^{23}$Na.
\end{abstract}
\maketitle
\section{Introduction}
Feshbach resonance\cite{Feshbach} has been observed in
$^{23}$Na\cite{Inouye}, $^{85}$Rb\cite{Cornish}, and a few other
alkali systems.  In these systems, two atoms can collide and fall
into a molecular state with a different spin.  The scattering
length can be tuned by changing the energy difference between the
atomic and molecular states through a magnetic field.  In the
$^{85}$Rb system, the scattering length was tuned to the negative
region to study the critical particle number of the
condensate\cite{Roberts}.  In the positive region, an unusual
processes of particle loss were observed\cite{Claussen}. Recently
oscillations between atomic and molecular states were
discovered\cite{Donley}.

One of the motivations to study Feshbach resonance is to create a
Bose system with a large and positive scattering length $a$. In
the traditional theory of a dilute Bose
gas\cite{Lee,Beliaev,Hugenholtz}, every physical quantity is
expressed as a power series in the gas parameter $\sqrt{8\pi
na^3}$, with higher order terms coming from high order
perturbations.  But the perturbation breaks down when the density
$n$ is of the order of $1/a^3$.  It is a challenge to find out
what happens to the system beyond the dilute region.  In recent
experiments\cite{Donley}, $na^3$ can be tuned as high as $0.5$.

However, even in the dilute region $na^3 \ll 1$, the system with a
Feshbach resonance is different from the system without a Feshbach
resonance because of the molecular states.  The observation of
oscillations between the atomic and molecular states\cite{Donley}
indicates macroscopic occupation of molecular states.  In fact, in
addition to photoassociation\cite{Wynar,Mckenzie}, Feshbach
resonance is considered as another method to study the mixture of
atomic and molecular condensates.

In Ref.\cite{Timmermans}, a two-channel model was constructed to study
Feshbach resonance, with the Hamiltonian given by
\begin{equation} \label{H}
H={\hbar^2 \over 2m} \nabla \psi^\dagger \cdot \nabla \psi+{\hbar^2 \over 4m}
\nabla \phi^\dagger \cdot \nabla \phi+{g_0 \over 2} {\psi^\dagger}^2 \psi^2
-\gamma (\phi^\dagger \psi^2+h.c.)+h\phi^\dagger \phi,
\end{equation}
where $\phi$ is the molecular field operator and $\psi$ is the atomic field
operator.  The magnetic detuning energy is given by $h$.  The coupling
constant of atomic fields is given by $g_0$.  The coupling constant between
atomic and molecular states is given by $\gamma$.  The sign of $\gamma$ is
chosen to be positive by gauge transformation.

Most studies\cite{Timmermans,Heinzen,Cusack,Calsamiglia} of this
model have been focused on condensate properties.  When multiple
modes were included, the condensates were found to be depleted
over time\cite{Goral}.  In Ref.\cite{Holland}, a Hartree-Fock
method was used to take into account the contribution from atom
pairs with opposite momentum, and the dynamics of the correlation
between the condensates and atom pairs was studied numerically.
However some properties of this model such as the complete
excitation spectrum and the decay process are still unclear.  In
this paper, we study these problems in the dilute region where the
scattering length $a$ is much bigger than the scattering length
far off the resonance $a_0$ but much smaller than the
interparticle distance, $a_0 \ll a \ll 1/ \root 3 \of n$.  A
simple saddle-point method is developed to handle this case.

\section{Condensate properties}
In the two-channel system, the numbers of atoms and molecules are not constants.
It is easier to work in the grand canonical ensemble with the grand
thermodynamical potential given by
\begin{equation}
F=H-\mu(\psi^\dagger \psi+2\phi^\dagger \phi),
\end{equation}
where $\mu$ is the chemical potential.
At zero temperature in the dilute region, both atoms and molecules condense.
The dominant contribution to the grand potential comes from the condensation,
given by
\begin{equation}
F_0={g_0 \over 2} |\psi_0|^4-\gamma (\phi_0^* \psi_0^2+c.c.)
+h|\phi_0|^2 -\mu(|\psi_0|^2+2 |\phi_0|^2),
\end{equation}
where $\psi_0$ and $\phi_0$ are the expectation values of the
atomic and molecular fields.  The values of $\psi_0$ and $\phi_0$
are determined from the saddle-point equations given by
\begin{eqnarray}
{\partial F_0 \over \psi_0} &=& (g_0|\psi_0|^2-2\gamma \phi_0-\mu)\psi_0^*=0,\\
{\partial F_0 \over \phi_0} &=& (h-2\mu)\phi_0^*-\gamma {\psi^*_0}^2=0.
\end{eqnarray}

In the region with positive scattering lengths, $h<0$ and $\mu>0$,
the nontrivial solution of the above equations is given by
\begin{eqnarray} \label{Density}
\psi_0^2 &=& {\mu \over g_{\rm eff}},\\
\phi_0 &=& {\gamma \psi_0^2 \over h-2\mu}, \label{phi}
\end{eqnarray}
where $g_{\rm eff}=g_0-2\gamma^2/(h-2\mu)$.
This solution is a minimum of the grand potential in the subspace of $\psi_0$
and a maximum in the subspace of $\phi_0$ because molecules have lower
energy than atoms.  It is dynamically stable in the condensate
space\cite{Timmermans}.

Compared to the traditional theory, the relation between the chemical potential
and the atomic condensate density given by Eq.(\ref{Density}) remains the same,
with $g_{\rm eff}$ now as the effective coupling constant of the atoms.
The scattering length is proportional to the effective coupling constant,
\begin{equation}\label{a}
a=mg_{\rm eff}/(4\pi\hbar^2)=a_0 (1-{\Delta \over h-2\mu}),
\end{equation}
where the scattering length in the absence of the resonance is given by
$a_0=mg_0/(4\pi\hbar^2)$, and the constant $\Delta$ is given by
$\Delta=2\gamma^2/g_0$.  The scattering length in the form of Eq.(\ref{a})
have been observed in the experiments\cite{Inouye,Donley}.

In the dilute region, the interparticle distance is much larger
than any other length scale.  Since the chemical potential $\mu$
is a function of the atomic condensate density, the absolute value
of the detuning energy $|h|$ is much larger than the chemical
potential $\mu$.  As a consequence, the molecular condensate
density is much less than the atomic condensate density,
\begin{equation}
{\phi_0^2\over\psi_0^2}\approx{\gamma^2\psi_0^2\over h^2}<{\mu\over 2|h|}\ll 1.
\end{equation}
These condensate properties are consistent with the findings in
Ref.\cite{Timmermans}.

As shown in the next section, due to the coupling between
atoms and molecules, the total molecule density is bigger than the molecular
condensate density by a factor of $1/Z$ as given in Eq.(\ref{Z}), because the
molecular eigenstates also contains atom pairs.  However, in the dilute region,
the total molecule density is still much smaller than the atomic condensate density.
The ratio of the molecule density to the atom density $n_m/n_a$ is approximately given
by
\begin{equation}
{n_m \over n_a} \approx{\gamma^2\psi_0^2\over Z h^2} \approx \pi na^3.
\end{equation}

\section{Excitations and quantum depletion}
In the dilute region, the system contains two types of
excitations, the molecular excitation and the atomic excitation.
The bare energy of a molecule is given by the detuning energy $h$.
The coupling to the atomic states strongly renormalizes the
molecule energy, and the molecular eigenstate becomes a
superposition of the bare molecular state and pairing states of
atoms.  This renormalization can be easily demonstrated in the
vacuum case where there are few particles.

In the vacuum, at zero temperature, the propagator of the
molecular field is given by
\begin{equation}\label{mp}
G_m({\bf q},\Omega)={1 \over
\Omega-{\epsilon_q \over 2}-h+2\mu-\Sigma_m({\bf q},\Omega)+i\delta},
\end{equation}
and the propagator of the atomic field is given by
\begin{equation}
G_a({\bf k},\omega)={1 \over \omega-\epsilon_k+\mu+i\delta},
\end{equation}
where $\Sigma_m({\bf q},\Omega)$ is the self energy of the molecular
propagator generated by the coupling between atoms and molecules,
\begin{eqnarray}
\Sigma_m({\bf q},\Omega) &=& 2i\gamma^2\int{d\omega \over 2\pi}
\int{d^3k \over (2\pi)^3}G_a({\bf k},\omega)G_a({\bf q}-{\bf k},\Omega-\omega)\\
&=& 2\gamma^2\int{d^3k \over (2\pi)^3}
{1 \over \Omega-\epsilon_k-\epsilon_{{\bf k}-{\bf q}}+2\mu+i\delta},\label{se}
\end{eqnarray}
and $\epsilon_k=\hbar^2 k^2/(2m)$.

There is an unphysical ultra-violet divergence in the integral in
Eq.(\ref{se}). As shown in many examples in the effective field
theory, the bare parameters are often divergent and the physical
parameters are always finite.  The divergence from the
perturbation can be removed by introducing a divergent counter
term in the Hamiltonian.  (In the context of a dilute Bose gas
without a Feshbach resonance, see Ref.\cite{Braaten}.) Here we
regularize the divergence by introducing a divergent part of the
bare molecule energy $-\gamma^2\int d^3k/[2(2\pi)^3\epsilon_k]$.
After the regularization, the self energy is given by
\begin{eqnarray}
\Sigma_m({\bf q},\Omega) &=& 2\gamma^2\int{d^3k \over (2\pi)^3}
({1 \over \Omega-\epsilon_k-\epsilon_{{\bf k}-{\bf q}}+2\mu+i\delta}+
{1 \over 2\epsilon_k})\\
&=& {\gamma^2 \over 2\pi} ({m \over \hbar^2})^{3 \over 2}
\sqrt{{\epsilon_q \over 2}-2\mu-\Omega}.
\end{eqnarray}

The pole of the molecular propagator in Eq.(\ref{mp}) is now given by
\begin{equation}
\Omega_q={\epsilon_q \over 2}-2\mu-{1 \over 4}\left(
\sqrt{{m^3 \gamma^4 \over 4 \pi^2 \hbar^6}-4h}-
-{m^{3 \over 2} \gamma^2 \over 2 \pi \hbar^3} \right)^2.
\end{equation}
Far off the resonance, the molecule energy is simply
given by $h+\epsilon_q/2-2\mu$;
near the resonance $m^3 \gamma^4/(\pi^2 \hbar^6) \gg 4|h|$,
the molecule energy is given by
\begin{equation}\label{meg}
\Omega_q \approx {\epsilon_q \over 2}-2\mu-
{4\pi^2 \hbar^6  h^2 \over m^3 \gamma^4}.
\end{equation}
Near the resonance and close to the pole, the propagator can be
approximated as
\begin{equation}
G_m({\bf q},\Omega)={Z \over \Omega-\Omega_q+i\delta},
\end{equation}
where the renormalization factor $Z$ is given by
\begin{equation}\label{Z}
Z={4\pi\hbar^3 \over m\gamma^2}\sqrt{|\Omega_0| \over m}.
\end{equation}
In this region, the scattering length is approximately given by
\begin{equation}
a \approx {\gamma^2 m \over 2\pi \hbar^2 |h|},
\end{equation}
and the molecular binding energy is approximately given by
\begin{equation}
\Omega_0 \approx -{4\pi^2 \hbar^6  h^2 \over m^3 \gamma^4}-2 \mu.
\end{equation}
The relation between the binding energy of molecules and the
scattering length is similar to that in the single-channel problem,
\begin{equation}\label{bind}
\Omega_0 \approx -{\hbar^2 \over m a^2}-2\mu,
\end{equation}
except that there is a mean-field shift by $2\mu$.

So far the results presented in this section
Eq.(\ref{meg}-\ref{bind}) are derived for the vacuum case, valid
near the resonance when the scattering length $a$ is much larger
than the scattering length off the resonance $a_0$. For a system
with many particles, these results are still valid as long as the
system is in the dilute region $a_0 \ll a \ll 1/ \root 3 \of n$,
because in this region the many-body contribution to the molecular
energy from the condensate is proportional to the particle density
$n$, much smaller than $|\Omega_0|$.  However, beyond the dilute
region when $na^3 \geq 1$, the many-body effect may be dominant
and the results for the dilute case may no longer be accurate.  In
a recent experiment\cite{Claussen2}, a shift to the molecular
energy larger than $2\mu$ was observed, which may be due to the
many-body effect.

If the molecule density is initially slightly away from the stationary value
due to a perturbation of the molecule energy in the action
$\alpha \phi^\dagger\phi$, the molecule density will oscillate.  The
oscillation can be calculated in the linear response theory. In the long-time
$t\rightarrow \infty$ limit, it is given by
\begin{eqnarray}
<\phi^\dagger(t)\phi(t)>-\phi_0^2 &\approx& -i\alpha\phi_0^2[<\phi_0(t)
\phi_0^\dagger(0)> -\phi_0^2] \\
&\approx& -\alpha\phi_0^2 \int{d\omega \over 2\pi} G_m(0,\omega)
e^{-i\omega t/\hbar}\\
&\approx& 2i{\alpha \Omega_0\over |h|}\phi_0^2 e^{-i\Omega_0t/\hbar}.
\end{eqnarray}
The oscillation frequency is simply given by the molecular binding
energy $\Omega_0$ because of the molecular condensation.  It is
similar to the single-channel problem where the phonon frequency
coincides with the energy of the single particle excitation.  In
the experiment\cite{Donley}, the oscillation frequency fits
perfectly with the molecule energy calculated from a microscopic
potential\cite{vanKempen}. This result was also obtained in the
context of photoassociation\cite{Mackie}, in the Hartree-Fock
approach\cite{Holland}, and in Ref.~\cite{Koehler}.

In the dilute region, most of the particles are in the atomic condensate.  The
long-wavelength and low-energy excitations are atomic excitations,
described by the gaussian fluctuation
\begin{equation}\label{quad}
\delta F_2=\sum_{\bf k}[(\epsilon_k-\mu+2g_0\psi_0^2-{2\gamma^2\psi_0^2 \over
h-2\mu})\psi_{\bf k}^\dagger \psi_{\bf k}+({g_0 \psi_0^2 \over 2}-\gamma \phi_0)
({\psi_{\bf k}^\dagger} \psi_{\bf -k}^\dagger+h.c.)].
\end{equation}
The molecular part of the fluctuation is not included because molecular
excitations have a gap $\Omega_0$ as shown above.  However molecular
excitations contribute a diagonal self-energy to the atomic propagator through the
coupling between atoms and molecules
\begin{equation}\label{sea}
\Sigma_a(k,\omega)=2\gamma^2\psi_0^2 G_m(k,\omega) \approx
-{2\gamma^2\psi_0 \over h-2\mu},
\end{equation}
which is already included in the r.h.s. of Eq.(\ref{quad}).

After diagonalizing Eq.(\ref{quad}), we obtain a linear dispersed mode
\begin{equation}
\delta F_2=C+\sum_{\bf k}E_k c_{\bf k}^\dagger c_{\bf k},
\end{equation}
where $c_{\bf k}=u_k \psi_{\bf k}+v_k \psi_{-{\bf k}}^\dagger$, $u_k^2={1
\over 2}({\epsilon_k+\mu \over 2E_k}+1)$ and $v_k^2=u_k^2-1$.
The excitation energy is given by
\begin{equation}\label{Ek}
E_k=\sqrt{\epsilon_k(\epsilon_k+2\mu)},
\end{equation}
and the energy of quantum depletion $C$ is given by
\begin{equation}\label{Eg}
{C \over V}={8 \over 15\pi^2}({m \mu\over \hbar^2})^{3 \over 2} \mu.
\end{equation}
The density of the quantum depletion can be easily calculated.
\begin{equation}\label{dn}
\delta n=\int{d^3k\over(2\pi)^3}v_k^2={1\over3\pi^2}
({m\mu\over\hbar^2})^{3\over2}
\end{equation}
These results are consistent with the first order
results in the traditional theory.  To this order,
$\mu=4\pi\hbar^2a/m$, any physical quantity can be expressed in terms
of the density of the atomic condensate $\psi_0^2$ and the scattering
length $a$ .

\section{The three-body recombination}
In the dilute region, the condensate state is only metastable
because occupation of molecular states can lower the system
energy.  Most of the decay is due to the three-body recombination.
In the two-channel system, the picture of this process is rather
simple.  One pair of atoms with opposite momentum in the quantum
depletion break up, and one of the them interacts with an atom in
the condensate to form a molecule leaving the other atom of the
original pair as an excited atom.  The energy and momentum can be
conserved in this process because molecules have negative
energies. Mathematically the decay comes from the perturbation of
the atom-molecule coupling term $-2\gamma\psi_0\phi_{\bf
k}^\dagger\psi_{\bf k}$ in the Hamiltonian.

The probability density of this three-body recombination
process can be calculated by Fermi's golden rule
\begin{eqnarray}
\Gamma &=& {2\pi \over \hbar}\int {d^3k \over (2\pi)^3}|<{\bf k}|
2\gamma\psi_0\phi_{\bf k}^\dagger\psi_{\bf k}|G>|^2 \delta(E_k+\Omega_k) \\
&=& {2\pi \over \hbar}\int {d^3k \over (2\pi)^3} 4 \gamma^2 \psi_0^2 v_k^2
Z \delta(E_k+\Omega_k),
\end{eqnarray}
where $|G>$ is the ground state of $\delta F_2$, $|{\bf k}>$ is the ground
state plus a molecule with wavevector ${\bf k}$ and an excited atom with
wavevector ${\bf -k}$. As defined earlier, $v_k$ is the coherence factor,
and $Z$ is the probability of finding a molecule in a molecular eigenstate
given in Eq.(\ref{Z}).
In the dilute region near resonance, the decay rate is approximately given by
\begin{eqnarray}
\Gamma &\approx& {2\pi \over \hbar}\int {d^3k \over (2\pi)^3}16\psi_0^2
{\mu^2 \over 4\epsilon_k^2} \sqrt{|\Omega_0|} ({m\over \hbar^2})^{3 \over 2}
\delta({3\epsilon_k \over 2}-{\hbar^2 \over m a^2})\\
&=& 32\sqrt{3}\pi^2{\hbar \over m}\psi_0^6 a^4. \label{Gama}
\end{eqnarray}

In Fig.(\ref{Fig}), the particle-loss rate for the $^{23}$Na
system is plotted and compared with the experimental
data\cite{Inouye}. In the region $2 a_0 \leq a \leq 10 a_0$ (906G
to 906.9G for the magnetic field), the theoretical result is
fairly close to the experimental data, although the theoretical
derivation so far is for the homogeneous case and the experimental
system is inhomogeneous. In the single-channel
case\cite{Fedichev}, the rate of the three-body recombination is
also proportional to $n^3a^4$, but the coefficient is about $3.9$,
about 140 times smaller than $32\sqrt{3}\pi^2$, and cannot be
fitted to the experimental result on $^{23}$Na.  The increase of
the coefficient reflects the fact that in the two-channel system
there is a virtual process allowing two atoms to collide and fall
into a molecular state, so the three-body recombination process is
greater.  However, in rubidium system, the experimental results
are rather intricate\cite{Claussen}. A decay rate with a weak
density dependence was observed, which seems to suggest that the
decay is dominated by a process independent of the particle
density.

In addition, the collision between atoms and molecules can also
provide a three-body decay but its decay rate is much
smaller\cite{amc}.  The decay of the condensate is consistent the
numerical result in Ref.~\cite{Goral} that when multiple modes are
included the condensate is depleted over time. The energy scale of
the decay rate is comparable to the chemical potential $\mu$ when
$\psi_0^2 a^3$ is of the order of $0.01$,
\begin{equation}
{\hbar \Gamma \over \psi_0^2 u} \approx 8\sqrt{3}\pi \psi_0^2 a^3.
\end{equation}
It indicates that the condensate may not be stable enough beyond
the dilute region.

\section{Conclusion}
In conclusion, we have obtained properties of a dilute Bose gas
with a Feshbach resonance at zero temperature.  The system
contains both atomic and molecular condensates but the atom
density is much higher.  The molecular excitation has a energy gap
and the atomic excitations are gapless. The low-energy properties
are consistent with the results of the traditional theory.
However, the condensed state is only metastable.  The coupling
between atomic and molecular states gives rise to a three-body
recombination process. The decay rate is proportional to $\psi_0^6
a^4$ as in the single-channel case, but the coefficient is much
bigger as observed in the Na experiment\cite{Inouye}.

It is tempting to go beyond the dilute region in the current
approach.  However, there are two expansion parameters in higher
orders terms neglected in this approach, $\psi_0a_0\sqrt{a}$ from
the atomic interaction and
$\psi_0m^2\gamma^2\sqrt{a}/(\hbar^4|h|)$ from the atom-molecule
coupling.  The latter is of the order of $\psi_0\sqrt{a^3}$ which
is much larger than unity near the resonance, and causes the
expansion to break down beyond the dilute region.  In the most
recent experiment\cite{Claussen2}, a shift in the molecular energy
was observed, which cannot be explained by any theories so far. It
seems to suggest that there is a non-trivial many-body effect
beyond the dilute region and more theoretical effort are needed.

We would like thank Tin-Lun Ho, David J. Thouless, and Ping Ao for
helpful discussions. LY was supported by NSF under Grant No. DMR
0201948 during the stay in University of Washington, and is
currently supported by SRF for ROCS, SEM.  ZHN is supported by
NSFC under Grant No. 10174003.

\begin{figure}
\caption{\label{Fig} The coefficients of the particle-loss rate
for Na system. $\dot N/N<n^2>$ is plotted versus the magnetic
field as in Fig.2 in Ref.\cite{Inouye}. The straight line is the
theoretical result computed from Eq.(\ref{Gama}); the squares and
triangles are the experimental data taken at ramp speeds
$0.13G/ms$ and $0.31G/ms$ of the magnetic field\cite{Inouye}. The
scattering length as a function of the magnetic field is computed
from Eq.(\ref{a}) with the parameters given in Ref.\cite{Inouye}.}
\end{figure}


\begin{thebibliography}{99}
\bibitem{amc}
The decay rate due to the atom-molecule collision is is proprotional
to $\psi_0^6 a_c^2/\gamma^4$, where $a_c$ is the scattering length of the
atom-molecule collision.  Near the resonance, $a \gg a_c$
and $a \gg {m^2 \gamma^2 \over \hbar^4}$, this decay rate is much
smaller than $\Gamma$.
\end{thebibliography}
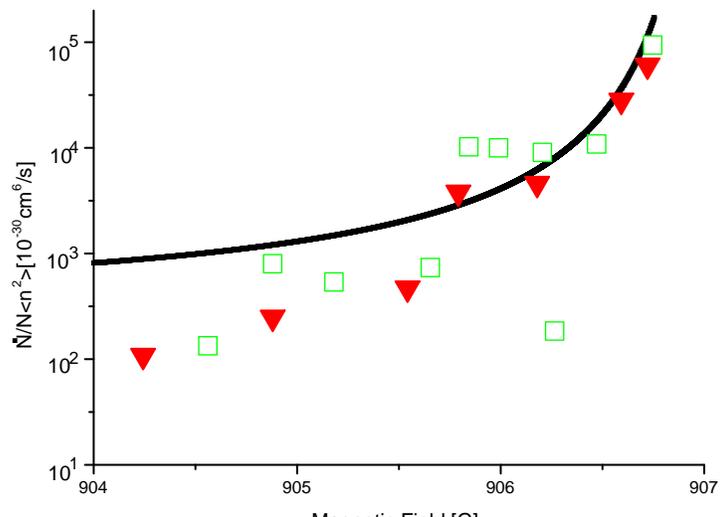
\end{document}